# An Assessment of the Consistency for Software Measurement Methods

Ramón Asensio Monge, Francisco Sanchis Marco and Fernando Torre Cervigón

**Abstract**- Consistency, defined as the requirement that a series of measurements of the same project carried out by different raters using the same method should produce similar results, is one of the most important aspects to be taken into account in the measurement methods of the software.

In spite of this, there is a widespread view that many measurement methods introduce an undesirable amount of subjectivity in the measurement process. This perception has made several organizations develop revisions of the standard methods whose main aim is to improve their consistency by introducing some suitable modifications of those aspects which are believed to introduce a greater degree of subjectivity.

Each revision of a method must be empirically evaluated to determine to what extent is the aim of improving its consistency achieved.

In this article we will define an homogeneous statistic intended to describe the consistency level of a method, and we will develop the statistical analysis which should be carried out in order to conclude whether or not a measurement method is more consistent than other one.

**Index Terms**- Software measurement, software metrics, measurement methods, consistency.

## 1 INTRODUCTION

The software is the main component of the corporation budget of any company. These organizations are aware of the importance of controlling the software costs and therefore, they analyse the profits of the resources assigned to its development and maintenance with the purpose of optimizing them. In order to do so, measurements and models using these magnitudes are needed [5].

Functional measurements are necessary for project management, which includes two principal functions: planning and control. Both of them require the ability to measure the software in an accurate and reliable way. Project management planning emphasizes the estimation of budgets. The development control requires measurement the project progress and permits to evaluate the efficiency of the tools used in the project development so as to improve the productivity.

Unfortunately, experience shows that these activities are not carried out accurately. The projects deviate from the budget, partly due to the inadequacy of the first estimations because of the subjectivity of the measurements [8] [6]. Consequently, a critical aspect in software management is to use reliable measurements, mostly in the first stages of the life-cycle. This will allow a more accurate estimation of the relationship between the product and the cost or the time required to develop it [4].

According to L. Ejiogu [3], the software metrics should have the following features:

- To be simple and easy to use. They should be easy to learn and their application should not involve a great effort or much time.

- To be convincing. The metrics must be in agreement with what our intuition suggest.

- To be reliable. Starting from the same information, several meter should obtain really similar results.

- To agree with the principles of the measurement theory. Mainly, as regards the performance of mathematical operations allowed by the corresponding scale type.

Some of the more commonly used metrics do not have one or several of these features.

## 2 CONSISTENCY OF A METHOD

Some researchers claim that there is a low inter-rater reliability in the application of a measurement method, that is several measurements of a given project by means of the same method give the same result. However, in practice, the results differ significantly. This notion, which both researchers and users of the measurement methods of the software support, has a negative impact on the confidence level of those methods and slows down their acceptance.

The possibilities either of using only a rater for measurement all the projects, or using for each project the average of the measurements by several raters are not valid on the grounds of flexibility and cost.

# 3 ANALYSIS OF THE CONSISTENCY LEVEL

The revisions on the software measurement methods, developed with the purpose of improving their consistency must be empirically evaluated so as to determine to what extent is the pursued goal fulfilled.

The aspects of each revised method (which for the sake of brevity will referred to only as a method from now on) that we will consider are those regarding:
  - inter-rater reliability
  - inter-method reliability

The definition of reliability used in this paper is that by Carmines and Zeller [1] who define it as: "the extent to which an experiment, test or any other type of procedure of measurement provides the same result in repeated tests. This tendency towards consistency found in repeated tests on the same phenomenon is what we refer to as reliability"

## 3.1 INTER-RATER RELIABILITY

To investigate the inter-rater reliability involves carrying out several measurements[i] by different raters with the same method, trying to find out whether or not such measurements produce similar results.

A formal planning of the experiments must be done, having into account the conception, design, preparation and fulfilment of the measurements. This planning has to comply with the principles of the design of experiments as regards their replication, randomization, local control, etc, in order to make the design as well as possible and to reduce experimental error. All these requirements as well as their involved cost explain the scarce research in this field and that the ones made are based upon a very reduced number of experiments [7] [4].

### 3.1.1 Influence of raters

Firstly, it is necessary to investigate the possible influence the raters may have in measurements. If $M_{ji}$ stands for the measurement by rater j of the project i and $M_{ki}$ the measurement by rater k of the project i with any method, then:

$$M_{ji} = X_i + \varepsilon_{ji}$$
$$M_{ki} = X_i + \varepsilon_{ki}$$

where $X_i$ is the unknown project real size, $\varepsilon_{ji}$ and $\varepsilon_{ki}$ have respectively $N(\tau_j ; \sigma^2)$ and $N(\tau_k ; \sigma^2)$ distributions, $d_{jki} = \varepsilon_{ji} - \varepsilon_{ki}$ has a normal distribution $N(\tau_j - \tau_k ; 2\sigma^2)$ and $d_{jki}/\sqrt{2} = (\varepsilon_{ji} - \varepsilon_{ki})/\sqrt{2}$ is normally distributed with mean equal to $(\tau_j - \tau_k)/\sqrt{2}$ and variance $\sigma^2$ )

When the null hypothesis $H_0$: $\tau_j - \tau_k = 0$ $\forall (j, k)$ is true, then
$d_{jki}/\sqrt{2} = (\varepsilon_{ji} - \varepsilon_{ki})/\sqrt{2}$ has a $N(0 ; \sigma^2)$ distribution.

Once the test has been carried out, if p-value > 0,05 then $H_0$ could be accepted. The acceptance of the null hypothesis implies admitting that the influence of all raters using the same method is the same.

The test will be applied for each method.

If the influence of the raters differs, it will be valuable to determine this influence and to remove it (or to eliminate the rater or raters who have obtained anomalous results). This usually signify, that those raters do not properly known the method.

### 3.1.2 Descriptive statistic of the consistency level

If the influence of raters coincides and if $M_{1Ai}$ denotes the first measurement with method A of a project i, and $M_{2Ai}$ denotes the second measurement with method A of a project i, the consistency level for method A will be given by the value of the descriptive statistic on the consistency level which is defined as the difference between measurements in an absolutely value divided by their average.

$$CA2i = \frac{|M_{1Ai} - M_{2Ai}|}{(M_{1Ai} + M_{2Ai})/2}$$

Similarly, we will proceed for the other methods.

### 3.1.3 Testing the equality of the consistency level of two methods

This process implies:
- To determine if the values of the statistic on the consistency level for each method are distributed in a normal way. In order to do so, the Kolmogorov-Smirnov test will be applied and if the p-value > 0,05 the normality could be accepted.
- To calculate the correlation coefficients of the consistency level statistics for each pair of methods. If the values of the consistency level have a normal distribution we will calculate Pearson's correlation coefficient, otherwise, that of Sperman, Kendall or any other coefficients will be applied.

If the p-value of the correlation coefficient is > 0,05, the samples of the values of the consistency level statistic can be assumed to be independent, otherwise, they are related.

**a.- Normal distribution and independent samples: Test of the equality of means.**

Let

$CA2_1\ CA2_2\ CA2_3\ \ldots\ldots\ldots\ldots\ CA2_{n1}$

$CB2_1\ CB2_2\ CB2_3\ \ldots\ldots\ldots\ldots\ CB2_{n2}$

be two independent samples of values of the statistic for a any pair of methods A and B. The mean of the first sample

$$\bar{x}_1 = \frac{\sum_{i=1}^{n1} CA2_i}{n1}$$

has a $N(\mu_1, \sigma_1/\sqrt{n1})$ distribution.

Equally,

$$\bar{x}_2 = \frac{\sum_{i=1}^{n2} CB2_i}{n2}$$

has a $N(\mu_2, \sigma_2/\sqrt{n2})$ distribution.

The difference $\bar{x}_1 - \bar{x}_2$

of means has a normal distribution with mean $\mu_1 - \mu_2$ and standard deviation:

$$\sqrt{\frac{\sigma_1^2}{n1} + \frac{\sigma_2^2}{n2}}$$

When the samples are large and the variances unknown, the statistic:

$$Z = \frac{\bar{x}_1 - \bar{x}_2}{\sqrt{\frac{s_1^2}{n1-1} + \frac{s_2^2}{n2-1}}}$$

has a N(0, 1) distribution if the null hypothesis $H_0 : \mu_1 = \mu_2$ is true.

$s_1^2\ s_2^2$ are the sample variances.

Let the significance level be $\alpha$; if

$$\frac{|\bar{x}_1 - \bar{x}_2|}{\sqrt{\frac{s_1^2}{n1-1} + \frac{s_2^2}{n2-1}}} \leq z_{\alpha/2}$$

the null hypothesis is accepted, otherwise it is rejected.

If the samples are small (n1 and n2 ≤ 30) the statistic

$$t = \frac{\bar{x}_1 - \bar{x}_2}{\sqrt{\left(\frac{n1 s_1^2 + n2 s_2^2}{n1 + n2 - 2}\right)\frac{n1 + n2}{n1 n2}}}$$

has a Student t distribution with n1+n2-2 degrees of freedom.

If

$$\frac{|\bar{x}_1 - \bar{x}_2|}{\sqrt{\left(\frac{n1 s_1^2 + n2 s_2^2}{n1 + n2 - 2}\right)\frac{n1 + n2}{n1 n2}}} \leq t_{\alpha/2, n1+n2-2}$$

then $H_0$ could be accepted at significance level $\alpha$. Otherwise, it should be rejected.

If the null hypothesis is rejected, a lower mean value of the consistency level statistic implies a greater consistency of the method.

If the null hypothesis is accepted, a test of equality of variances will be made.

**b.- Normal populations and independent samples: Test of equality in of variances.**

Let $CA2_1\ CA2_2\ CA2_3\ \ldots\ldots\ldots\ CA2_{n1}$ and $CB2_1\ CB2_2\ CB2_3\ \ldots\ldots\ldots\ CB2_{n2}$ denote, respectively, independent samples from the two independent distributions the consistency level statistic, having, respectively the probability density functions: N($\mu_1$, $\sigma_1$) and N($\mu_2$, $\sigma_2$). Then

$$F = \frac{n1(n2-1)s_1^2}{n2(n1-1)s_2^2}$$

has a Snedecor F distribution with (n1−1), (n2−1) degrees of freedom.

If F is in the confidence interval

$$\left(F_{1-\alpha/2, n1-1, n2-1},\ F_{\alpha/2, n1-1, n2-1}\right)$$

the null hypothesis could be accepted.

The acceptance of the null hypothesis implies that we cannot conclude that a method is more reliable than the other one.

If the null hypothesis is rejected, a lower value of the variance of the consistence level statistic implies a greater consistency of the corresponding method.

**c.- Normal distribution and related samples: Test of the equality of two means.**

The distribution of the difference of means of the two related samples has the same characteristics that of independent samples, except the standard error, which is smaller.

In this case, the statistic:

$$Z = \frac{\overline{x}_1 - \overline{x}_2}{\sqrt{\frac{s_1^2}{n1-1} + \frac{s_2^2}{n2-1} - 2r_{12}\frac{s_1}{\sqrt{n1-1}}\frac{s_2}{\sqrt{n2-1}}}}$$

where $r_{12}$ is the correlation coefficient, has a N(0, 1) distribution if the null hypothesis $H_0: \mu_1 = \mu_2$ is true.

Let $\alpha$ be the significance level, if:

$$\frac{|\overline{x}_1 - \overline{x}_2|}{\sqrt{\frac{s_1^2}{n1-1} + \frac{s_2^2}{n2-1} - 2r_{12}\frac{s_1}{\sqrt{n1-1}}\frac{s_2}{\sqrt{n2-1}}}} \leq z_{\alpha/2}$$

we can accepted the null hypothesis.
If the hypothesis is accepted, a test of the equality of variances will be made.
If the null hypothesis is rejected, a lower mean value of the consistency level statistic implies a greater consistency of the method.

**d.- Normal distributions and related samples: Test of the equality of variances.**

In the present case, the statistic:

$$-\left[n - 1 - \frac{p(p+1)^2(2p-3)}{6(p-1)(p^2+p-4)}\right]$$

$$\ln\left[\frac{|\hat{s}|}{(\hat{\sigma}^2)(1-\hat{\rho})^{p-1}(1+(p-1)\hat{\rho})}\right]$$

has a chi-square $\chi^2$ distribution with ½ p(p+1)-2 degrees of freedom when the null hypothesis $H_0: \sigma_1^2 = \sigma_2^2$ is true. In this study, p is the dimension of the random vector and equals 2, and , and n1=n2=n is the number of projects measured with methods A and B.
Let

$$\hat{s}_{11} = \frac{1}{n-1}\sum_{i=1}^{n}(CA2_i - \overline{x}_1)^2$$

$$\hat{s}_{22} = \frac{1}{n-1}\sum_{i=1}^{n}(CB2_i - \overline{x}_2)^2$$

$$\hat{s}_{12} = \frac{1}{n-1}\sum_{i=1}^{n}(CA2_i - \overline{x}_1)(CB2_i - \overline{x}_2) = \hat{s}_{21}$$

$$\hat{s} = \begin{pmatrix} \hat{s}_{11} & \hat{s}_{12} \\ \hat{s}_{21} & \hat{s}_{22} \end{pmatrix}$$

be the matrix of variances-covariances, and let

$|\hat{s}|$ be the determinant of the matrix of variances co-variances, and

$$\hat{\sigma}^2 = \frac{\hat{s}_{11} + \hat{s}_{22}}{2}$$

$$\hat{\rho} = \frac{\hat{s}_{12}}{\hat{\sigma}^2}$$

Given a significance level $\alpha$, we calculate the critical value $K_\alpha$.
If the statistic value is greater than $K_\alpha$ the null hypothesis of equality of variances is rejected. In this case, a lower value of the variance of the consistency level implies a greater consistency of the corresponding method.

**e.- Non-normal distributions: Test of the equality of two independent distributions. Small samples.**

For non-normal distributions, the nonparametric test of Mann – Whitney will be applied. In order to it, we proceed as follows:

- We rank jointly the two samples, i.e.:

CA2$_1$  CA2$_2$  CA2$_3$ .................... CA2$_{n1}$

CB2$_1$  CB2$_2$  CB2$_3$ .................... CB2$_{n2}$

- The rank or order 1 is assigned to the algebraically smallest punctuation. When two data from the samples have the same value, the average value of the ranks or orders is assigned to it.

- Calculate:

- $R_A$ sum of ranks from sample A.
- $R_B$ sum of ranks from sample B.
- $N_A$ number of elements in sample A
- $N_B$ number of elements in sample B.

- Determine:

$$U_A = N_A N_B + \frac{N_A(N_A+1)}{2} - R_A$$

$$U_B = N_A N_B + \frac{N_B(N_B+1)}{2} - R_B$$

$$U_A + U_B = N_A N_B$$

- We take the smallest of $U_A$ or $U_B$ and we compare it with the critical value of U, which is determined in the Mann – Whitney table in function of $N_A$ and $N_B$ and the specific significance level $\alpha$.

- If $U_A$ (o $U_B$) < U the null hypothesis should be rejected.

If the null hypothesis is rejected, a lower value of the mean of the consistency level statistic, implies a greater consistency of the corresponding method.

**f.- Non-normal distributions: Test of the equality of two independent distributions. Large samples.**

When $N_A$ and $N_B$ increase their size, Mann-Whitney proved that U has approximately a normal distribution, with:

mean $$\mu_U = \frac{N_A N_B}{2}$$

and standard deviation

$$\sigma_U = \sqrt{\frac{N_A N_B (N_A + N_B + 1)}{12}}$$

Let $\alpha$ be a significance level and $N_A$ and $N_B$ > 10; if

$$\frac{\left| U - \frac{N_A N_B}{2} \right|}{\sqrt{\frac{N_A N_B (N_A + N_B + 1)}{12}}} > Z_{\alpha/2}$$

where U is the minimum of de $U_A$ or $U_B$, the difference of the two samples is considered to be significant. In this case, a lower value of the mean of the consistency level statistic implies a greater consistency in the corresponding method.

**g.- Non-normal distributions: Test of the equality of two related distributions. Small samples.**

For small samples whose data are related, we apply the Wilcoxon T test, which is similar to the sign test, but more powerful and efficient, since it takes into account, not only the signs but also the magnitudes of the difference among data.

The Wilcoxon T test includes the following steps:
- Computing the difference between data of both samples, having into account the sign.
- Ranking these differences giving rank 1 to the smallest one, rank 2 to the following one, and so on. This order is carried out disregarding the signs. If there are two equal differences, the rank assessed to them is the average of both. On the other hand, if the difference is zero, it will be disregarded.
- $T_P$ is the sum of the ranks corresponding to positive differences.
- $T_N$ is the sum of the ranks corresponding to negative differences. It should be true that

$$T_P + T_N = \frac{n(n+1)}{2}$$

- Let $\alpha$ be a significance level; the critical value of T is determined for a sample of size n, in the corresponding Wilcoxon table.
- If $T_P$ or $T_N$ is lower than equal to the critical value of T, the difference of the two samples is significant. In such case, a lower of the mean of the consistency level statistic implies a greater consistency in the corresponding method.

**h.-Non-normal distributions: Test of the equality of two related distributions. Large samples.**

When the samples are large, that is when n is higher than 30, an approximation to the normal distribution can be used, with:

mean = $\overline{T} = \dfrac{n(n+1)}{4}$

standard deviation = $\sigma_T = \sqrt{\dfrac{n(n+1)(2n+1)}{24}}$

Let $\alpha$ be the significance level; if:

$$\dfrac{\left|T - \dfrac{n(n+1)}{4}\right|}{\sqrt{\dfrac{n(n+1)(2n+1)}{24}}} > Z_{\alpha/2}$$

Where, T is the minimum of $T_P$ and $T_N$, the difference of the two samples is significant.
In this case, a lower value of the mean of the consistency level statistic implies a greater consistency in the corresponding method.

## 3.2 INTER–METHOD RELIABILITY

The matter is to research if groups of raters using two different methods produce similar results.

### 3.2.1 Eliminate the possible influence of raters

In order to eliminate the possible influence of raters, we consider for each pair of methods A and B the new variable dab, which is given by the arithmetical mean of the differences between the measurements by each project with each method:

d1ab = $M_{1A}$ - $M_{1B}$

d2ab = $M_{2A}$ - $M_{2B}$

dab = ( d1ab + d2ab ) / 2

A and B are any two methods.

### 3.2.2 Testing the equality of the measurements given by two methods

We will consider the new variable dab.
Let us state the null hypothesis: $H_0$: dab = 0.
A test of Kolmogorov-Smirnov will be applied in order to determine if dab has a normal distribution.
If dab is normal, then:

$$t = \dfrac{\overline{dab}}{\dfrac{s_{dab}}{\sqrt{n-1}}}$$

has a Student t distribution with n-1 degrees of freedom.
Let $\alpha$ be the significance level; if:

$$\dfrac{\left|\overline{dab}\right|}{\dfrac{s_{dab}}{\sqrt{n-1}}} > t_{\alpha/2, n-1}$$

the null hypothesis should be rejected. To reject the null hypothesis implies significant differences between the measurements given by methods A and B are detected.

### 3.2.3 Fitting a regression function

If there are significant differences, we lose the advantages of the use of standard methods, which allow us the use the empirical models of cost estimation. In this case, it is possible to make a suitable cost estimation only after having applied the methods for a certain period of time.

If the correlation between methods is strong, it is possible to fit a regression function which allows us to obtain the value of the normalized data starting from the value of the measurement given by the revision on the method in use.

## 4 CONCLUSIONS

Consistency is a fundamental feature in the software metrics. In order to improve consistency several organizations have developed revisions of the standard methods, that is, modifications to some aspects of the method which have an influence on the consistency, however, without modifying its application domain.

In this work we develop the steps to follow in order to evaluate to which extent are these revisions significantly more reliable than the corresponding standard method.

To do so it is necessary to carry out a number of measurements by different raters of the same projects with the standard method and with a revision on the standard. From these measurements we will start by determining the possible influence of raters on measurements. If such influence is the same, we will define an homogeneous descriptive statistic which is given by the difference of measurements of each project, in an absolute value, divided by their average.

Depending on whether the distributions are normal or not and whether the samples are independent or related we show the corresponding statistic in relation to the test to carry out, their distribution and the critical value which will lead us to either accept or reject the formulated hypothesis.

The acceptance or rejection of the hypothesis, along with the value of the mean and/or the variance of the consistence level statistic, will allow us to decide if one method is more reliable than another.

What has been developed in this research has been applied to determine if successive revisions on the IFPUG method for the measurement of the functional size of software, developed in order to improve its consistency, would comply with the followed aim. Thirty projects have been measured each of them by two different raters using the IFPUG method and these same projects have also been measured using a revision on the same method. The results of such research will be matter for a new near future publication.

**REFERENCES**

[1] Carmines E.G. and Zeller R.A., "Reliability and Validity Assessment", Sage Publications, Beverly Hills, 1979

[2] Common Software Measurement International Consortium (COSMIC), "Measurement Manual", Version 2.0 October 1999

[3] Ejiogu L., "Software Engineering with Formal Metrics", QED Publishing 1991

[4] Kemerer C.F., "Reliability of function point measurement: a field experiment", Communications of the ACM, Volume 36, pp. 85-97, 1993

[5] St-Pierre D., Maya M., Abran A., Desharnais J.M. and Bourque P., "Full Function Poins: Counting Practices Manual", Software Engineering Management Research Laboratory Universite du Quebec a Montreal (UQAM) and Software Engineering Laboratory in Applied Metrics (SELAM), September 1997

[6] Rubin H.A., "A comparison of cost estimation tools", Proceedings of the International Conference on Software Engineering, IEEE Computer Society Press, 1985

[7] Rudolph E.E., "Productivity in Computer Application Development", University of Auckland, Dept. of Management Studies, New Zealand 1983

[8] Sanchís Marco F., "Proyectos Informáticos", Departamento de Publicaciones de la Escuela Universitaria de Informática, Universidad Politécnica de Madrid, 1998

. Ramón Asensio Monge is with the Department of Computer Science, Oviedo University, Spain.
E-mail: asensio@correo.uniovi.es

. Francisco Sanchis Marco is with the Department of EIO, Polytechnical University of Madrid. Spain.
E-mail: fsanchis@eui.upm.es

. Fernando Torre Cervigón is with the Department of Computer Science, Oviedo University, Spain.
E-mail: torre@lsi.uniovi.es

[i] The research has been developed on two measurements by project carried out by different meters with the same method. However, it can be easily applied to n measurements.